\immediate\write18{makeindex \jobname.nlo -s nomencl.ist -o \jobname.nls}
\documentclass[preprint, 5p]{elsarticle}
\usepackage{microtype}
\usepackage{amsmath}
\usepackage{textcomp}
\usepackage{booktabs}
\usepackage{array}
\usepackage[colorlinks]{hyperref}
\usepackage[capitalise,nameinlink]{cleveref}
\usepackage{siunitx}
\usepackage{gensymb}
\usepackage{graphicx}
\usepackage{subcaption}
\usepackage{framed}
\usepackage{multicol}
\usepackage{multirow}
\usepackage{nomencl}
\usepackage[labelfont=bf,justification=raggedright,singlelinecheck=true]{caption}
\usepackage{bm}
\usepackage{fancyhdr}

\makenomenclature
\renewcommand*\nompreamble{\begin{multicols}{2}}
\renewcommand*\nompostamble{\end{multicols}}

\biboptions{square,comma,numbers,sort&compress}

\begin{document}

\begin{frontmatter}

\title{In-Orbit Aerodynamic Coefficient Measurements using SOAR (Satellite for Orbital Aerodynamics Research)}

\author[UoM]{N.H.~Crisp}\corref{cor1}
\ead{nicholas.crisp@manchester.ac.uk}

\author[UoM]{P.C.E.~Roberts}
\author[UoM]{S.~Livadiotti}
\author[UoM]{A.~Macario~Rojas}
\author[UoM]{V.T.A.~Oiko}
\author[UoM]{S.~Edmondson}
\author[UoM]{S.J.~Haigh}
\author[UoM]{B.E.A.~Holmes}
\author[UoM]{L.A.~Sinpetru}
\author[UoM]{K.L.~Smith}

\author[Deimos]{J.~Becedas}
\author[Deimos]{R.M.~Dom\'{i}nguez}
\author[Deimos]{V.~Sulliotti-Linner}

\author[Gomspace]{S.~Christensen}
\author[Gomspace]{J.~Nielsen}
\author[Gomspace]{M.~Bisgaard}

\author[Stuttgart]{Y-A.~Chan}
\author[Stuttgart]{S.~Fasoulas}
\author[Stuttgart]{G.H.~Herdrich}
\author[Stuttgart]{F.~Romano}
\author[Stuttgart]{C.~Traub}

\author[UPC]{D.~Garc\'{i}a-Almi\~{n}ana}
\author[UPC]{S.~Rodr\'{i}guez-Donaire}
\author[UPC]{M.~Sureda}

\author[MSSL]{D.~Kataria}

\author[Euroconsult]{B.~Belkouchi}
\author[Euroconsult]{A.~Conte}
\author[Euroconsult]{S.~Seminari}
\author[Euroconsult]{R.~Villain}

\address[UoM]{The University of Manchester, Oxford Rd, Manchester, M13~9PL, United Kingdom}
\address[Deimos]{Elecnor Deimos Satellite Systems, Calle Francia 9, 13500 Puertollano, Spain}
\address[Gomspace]{GomSpace A/S, Langagervej 6, 9220 Aalborg East, Denmark}
\address[Stuttgart]{Institute of Space Systems (IRS), University of Stuttgart, Pfaffenwaldring 29, 70569 Stuttgart, Germany}
\address[UPC]{UPC-BarcelonaTECH, Carrer de Colom 11, 08222 Terrassa, Barcelona, Spain}
\address[MSSL]{Mullard Space Science Laboratory, University College London, Holmbury St. Mary, Dorking, RH5 6NT, United Kingdom}
\address[Euroconsult]{Euroconsult, 86 Boulevard de Sébastopol, 75003 Paris, France}

\journal{Acta Astronautica}

\begin{abstract}
The Satellite for Orbital Aerodynamics Research (SOAR) is a CubeSat mission, due to be launched in 2021, to investigate the interaction between different materials and the atmospheric flow regime in very low Earth orbits (VLEO). Improving knowledge of the gas-surface interactions at these altitudes and identification of novel materials that can minimise drag or improve aerodynamic control are important for the design of future spacecraft that can operate in lower altitude orbits. Such satellites may be smaller and cheaper to develop or can provide improved Earth observation data or communications link-budgets and latency. In order to achieve these objectives, SOAR features two payloads: i) a set of steerable fins which provide the ability to expose different materials or surface finishes to the oncoming flow with varying angle of incidence whilst also providing variable geometry to investigate aerostability and aerodynamic control; and ii) an ion and neutral mass spectrometer with time-of-flight capability which enables accurate measurement of the in-situ flow composition, density, velocity. Using precise orbit and attitude determination information and the measured atmospheric flow characteristics the forces and torques experienced by the satellite in orbit can be studied and estimates of the aerodynamic coefficients calculated. This paper presents the scientific concept and design of the SOAR mission. The methodology for recovery of the aerodynamic coefficients from the measured orbit, attitude, and in-situ atmospheric data using a least-squares orbit determination and free-parameter fitting process is described and the experimental uncertainty of the resolved aerodynamic coefficients is estimated. The presented results indicate that the combination of the satellite design and experimental methodology are capable of clearly illustrating the variation of drag and lift coefficient for differing surface incidence angle. The lowest uncertainties for the drag coefficient measurement are found at approximately \SI{300}{\kilo\meter}, whilst the measurement of lift coefficient improves for reducing orbital altitude to \SI{200}{\kilo\meter}.
\end{abstract}

\begin{keyword}
Orbital Aerodynamics; Drag and Lift Coefficient; Gas-Surface Interactions; Thermospheric Wind; CubeSat.
\end{keyword}

\end{frontmatter}

\begin{table*}[!t]   
	\begin{framed}
		\nomenclature[E]{$F$}{Force}
		\nomenclature[E]{$T$}{Torque}
		\nomenclature[E]{$m$}{Mass}
		\nomenclature[E]{$s$}{Molecular speed ratio}
		\nomenclature[E]{$I$}{Moment of inertia}
		\nomenclature[E]{$C_T$}{Torque coefficient}
		\nomenclature[E]{$C_F$}{Force coefficient}
		\nomenclature[E]{$T_w$}{Surface (wall) temperature}
		\nomenclature[E]{$T_\infty$}{Free-stream temperature}
		\nomenclature[G]{$\rho$}{Atmospheric density}
		\nomenclature[E]{$v_{\mathrm{rel}}$}{Relative atmospheric flow velocity}
		\nomenclature[E]{$A_{\mathrm{ref}}$}{Reference area}
		\nomenclature[E]{$A_T$}{Total Surface Area}
		\nomenclature[E]{$l_{\mathrm{ref}}$}{Reference length}
		\nomenclature[G]{$\alpha$}{Thermal (energy) accommodation coefficient}
		\nomenclature[G]{$\alpha_n$}{Normal energy accommodation coefficient}
		\nomenclature[G]{$\sigma_t$}{Tangential momentum accommodation coefficient}
		\nomenclature[F]{$\ddot{x}$}{Linear acceleration}
		\nomenclature[G]{$\ddot{\theta}$}{Rotational acceleration}
	\printnomenclature
	\end{framed}
	\end{table*}

\section{Introduction}
The Satellite for Orbital Aerodynamics Research (SOAR) is a scientific CubeSat mission due to be launched in 2021 and designed to investigate the interactions between the atmospheric flow regime in very low Earth orbits (VLEO) and different materials. Secondary objectives of the SOAR mission are to provide new in-situ measurements of the atmospheric density and composition and variation of the thermospheric wind velocity over the range of altitudes below approximately \SI{400}{\kilo\meter}. SOAR will also demonstrate novel attitude and orbit control manoeuvres using the aerodynamic forces and torques that can be generated at these altitudes. 

The SOAR mission is a key component of the Horizon 2020 funded DISCOVERER project \cite{Roberts2017,Roberts2019} that aims to radically redesign Earth observation satellites for sustained operation at significantly lower altitudes. The experiments performed by SOAR aim to improve knowledge and understanding of the gas-surface interactions (GSIs) at VLEO altitudes and provide valuable validation data for ground-based experiments on materials and GSIs which will be performed in the ROAR (Rarefied Orbital Aerodynamics Research) facility at The University of Manchester. The ROAR Facility is a unique experimental set-up that is designed to identify novel materials for satellite applications with a focus on improved aerodynamic properties and atomic oxygen (AO) resistance. The facility is principally comprised of a ultra-high vacuum (UHV) environment, an AO source capable of providing representative orbital velocities and surface interactions, and a sensor suite including ion and neutral mass spectrometers (INMS) which enable measurement and characterisation of the incident and re-emitted gas-flow on sample materials \cite{Oiko2018,Oiko2019}.

Improvements in the knowledge and understanding of the GSIs and identification of novel materials that can reduce atmospheric drag, improve aerodynamic control capability, or increase aerodynamic intake efficiencies are important steps in enabling the sustained operation of spacecraft at lower orbital altitudes. This reduction in orbital altitude has been linked to numerous benefits, for example reduced debris collision risk, a more favourable radiation environment, and aerodynamics-assisted end-of-life disposal. The opportunity to incorporate novel technologies such as atmosphere-electric propulsion (ABEP) and aerodynamic attitude and orbit control is also presented. For Earth observation applications, lower altitude orbits offer the possibility of smaller and less expensive platforms, leading to cheaper data products, or alternatively higher resolution imagery, both with a wide range of potential commercial, environmental, and societal impact \cite{Crisp2020}. Communications satellites may correspondingly benefit in their design from improved link-budgets, lower latency, and increased frequency re-use \cite{Gavish1998}.

\subsection{Gas-Surface Interactions in Very Low Earth Orbit}
\begin{figure}[tbh]
	\centering
	\includegraphics[width=85mm]{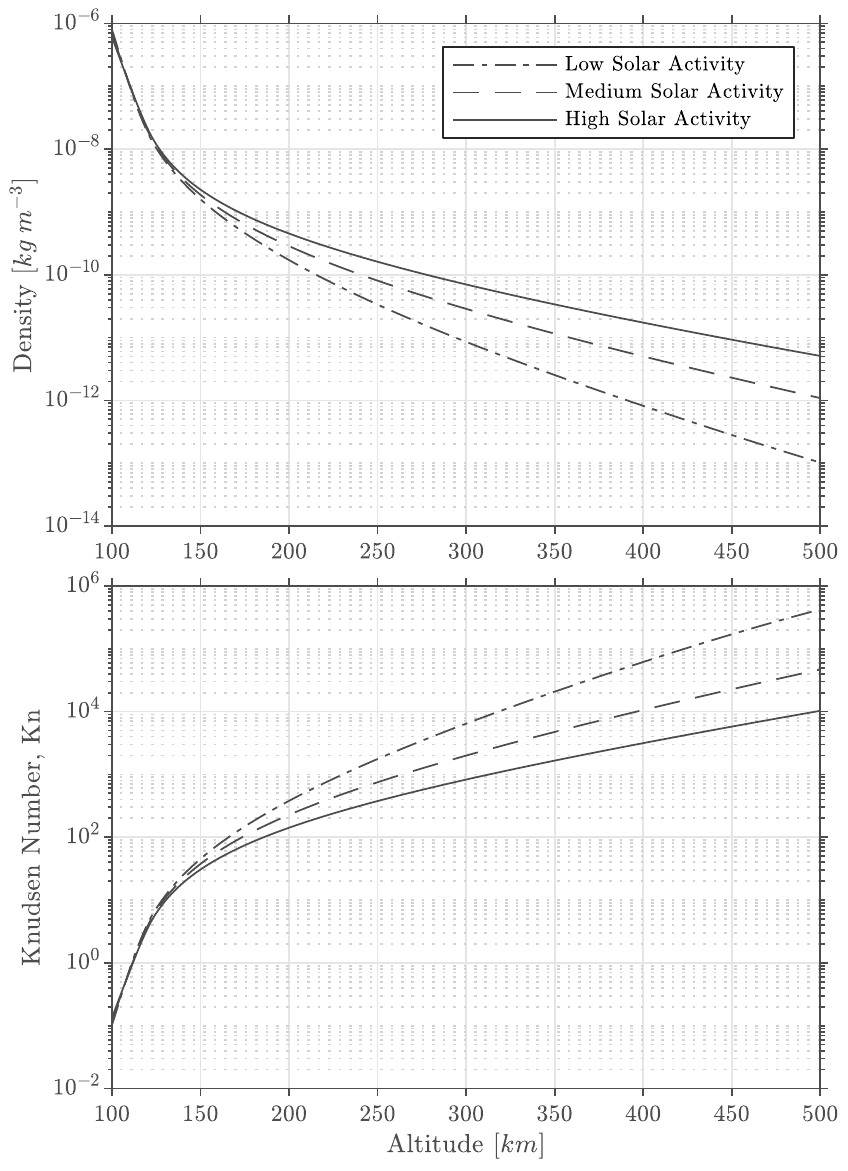}
	\caption{Variation of density (top) and free-stream Knudsen number (bottom) with altitude for different levels of solar activity, assuming a characteristic length of \SI{1}{\meter} and atmospheric parameters calculated using the NRLMSISE-00 model \cite{Picone2002}. Average kinetic diameter is weighted by the number density of the atmospheric species at each altitude (values for AO and N assumed to be conservatively equivalent to $\mathrm{N_2}$).}
	\label{F:Knudsen}
	\end{figure}

The upper bound of the VLEO range can be broadly defined as the altitude below which the atmosphere begins to have a significant effect on the orbital and attitude dynamics of a spacecraft and is typically defined at \SI{450}{\kilo\meter} altitude. However, this definition is somewhat fuzzy as in reality the the atmospheric density can vary considerably at this altitude (as shown in \cref{F:Knudsen}) principally as a result of the expansion and contraction of the atmosphere with the different diurnal, seasonal, and solar cycles.

In VLEO the atmosphere is significantly less dense than at the ground or conventional flight altitudes and is considered to be rarefied such that the mechanics of continuum flow regimes can no longer be applied. The non-dimensional Knudsen number can be used to classify different flow-regimes and is defined as the ratio between the mean free path (the average distance between successive gas particle to gas particle or gas particle to surface collisions) in a flow and a characteristic physical length (e.g. the length of a body in that flow). When the Knudsen number is high (i.e. $Kn \gg 10$) the gas-surface interactions along the length of a body are of much greater significance than any gas particle to gas particle interactions, including those with reflected particles \cite{Sentman1961}. This regime is termed free-molecular flow (FMF). The variation of the Knudsen number with altitude is given in \cref{F:Knudsen}. The lower bound of the VLEO range can be defined as the flow enters the more complex transitional regime ($Kn < 10$), and the conditions of free-molecular flow cannot be assumed. This is shown to occur for altitudes below approximately \SI{130}{\kilo\meter} altitude.

In the FMF regime, the forces which act on a body can be determined by simply considering the interaction between the incident molecules and satellite surfaces, and the subsequent angular distribution and velocity of the re-emitted or reflected particles. It has been observed that these GSIs, and the associated momentum and energy transfer, are dependent on surface roughness and cleanliness (particularly related to altitude-dependent AO adsorption), surface composition and lattice structure, surface temperature, gas composition, and the incident particle temperature, velocity, and incidence angle \cite{Moe1998,Bowman2005,Sutton2009a}. The presence of ionised thruster plumes may also affect the local flow conditions and therefore the aerodynamic forces produced \cite{Andrews2020}.

Models for these GSIs have been developed to enable estimation and determination of the aerodynamic forces which act on surfaces in these conditions. These models are used for the purpose of orbit and attitude simulation, spacecraft design and modelling, and in the development of atmospheric density and thermospheric wind models from on-orbit observations \cite{Doornbos2012}. Popular GSI models used in the field of orbit aerodynamics include those of \citet{Schaaf1958,Schamberg1959,Sentman1961,Gaposchkin1994,Storch2002}. Comparison and review of these models is provided by \citet{MostazaPrieto2014,Livadiotti2020}. 

In general, the re-emitted or reflected particle distribution is described as diffuse or specular with some models using combinations of these definitions. Coefficients that define the range of energy or momentum accommodation at the surface are typically used to characterise the GSI performance given an assumed re-emission distribution and therefore the forces experienced by the surface.

\begin{figure}
	\centering
	\includegraphics[width=85mm]{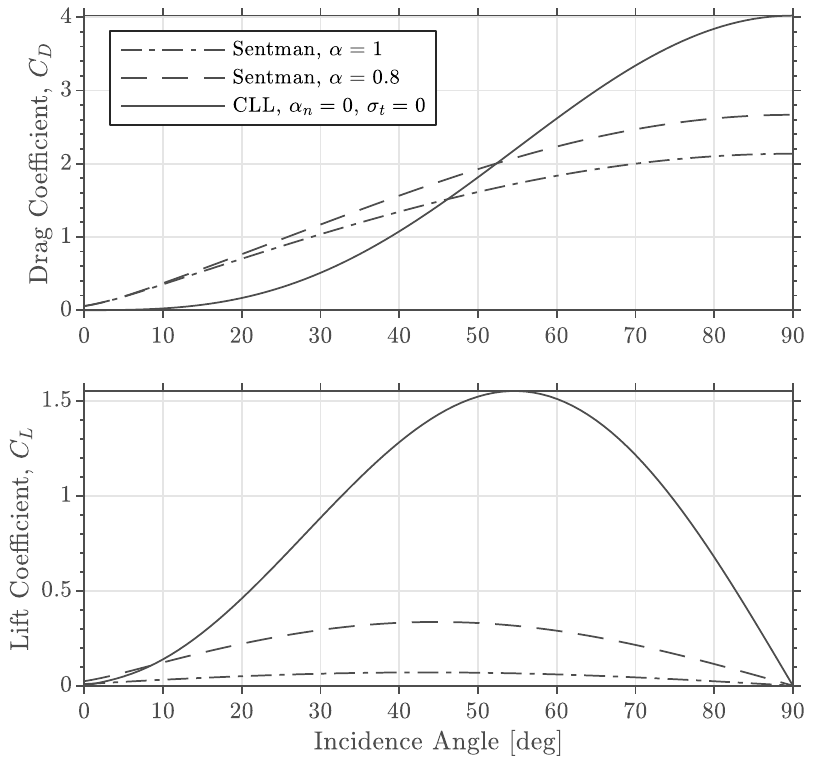}
	\caption{Variation of drag and lift force coefficients of a single-sided flat-plate of area \SI{1}{\meter\squared} under different GSI models and inputs assumptions for FMF conditions ($s = 10$, $T_w = \SI{300}{\kelvin}$, $T_\infty = \SI{600}{\kelvin}$). Aerodynamic coefficients are referred to the projected (cross-sectional) area with respect to the oncoming flow.}
	\label{F:GSI_CD_CL} 
	\end{figure}

\cref{F:GSI_CD_CL} demonstrates the effect of different GSI model assumptions and given parameters on the drag and lift force coefficients for a flat-panel surface. Sentman's model \cite{Sentman1961} assumes diffuse re-emission of particles, whilst modified analytical equations (from \citet{Schaaf1958}) based on the Cercignani-Lampis-Lord (CLL) model \cite{Cercignani1971,Lord1991,Lord1995} as proposed by \citet{Walker2014} can be used to represent specular reflections. For diffuse re-emission but reducing energy accommodation the drag force increases as the incidence approaches normal to the flow. Meanwhile, the lift force increases modestly for inclined surfaces to a maximum at approximately \ang{45}. However, if quasi-specular re-emission or specular reflection properties are exhibited the drag can be significantly reduced for shallow incidence angles (\textless\ang{45}) and will increase as the incidence approaches normal to the flow. Lift force generation can also be increased significantly.

Studies of in-orbit GSI performance have shown that materials commonly utilised on spacecraft have exhibited primarily diffuse re-emission properties with high energy accommodation ($\alpha \approx \;$\numrange{0.8}{1.0}), particularly in low altitude orbits where surface contamination (principally by adsorbed atomic oxygen) is high \cite{Moe1993,Moe1998}. The prevalence of energetic and highly-reactive atomic oxygen in low altitude orbits also introduces the issue of material erosion \cite{Reddy1995,Banks2004,Samwel2014} that can further increase accommodation and therefore result in diffuse re-emission. However, evidence of increasing quasi-specular re-emission behaviour has been observed for materials on spacecraft in higher altitude orbits (\SIrange{800}{1000}{\kilo\meter}) \cite{Harrison1996} where surface contamination is lower and in elliptical orbits where the incident kinetic energy near perigee is greater \cite{Moe1993}. Ground-based molecular beam experiments have also demonstrated such quasi-specular qualities for clean materials under UHV conditions and at energies approaching that of orbit velocity \cite{Murray2017}.

For the purpose of improving aerodynamic performance in the VLEO regime, quasi-specularly reflecting materials in combination with appropriate satellite geometric design, would provide the ability to reduce aerodynamic drag and therefore increase orbital lifetime or reduce the requirements for drag compensating propulsion systems. Alternatively, using the increased drag generated at high-incidence angles, enhanced aerodynamics-based deorbit devices could be conceived. The capability to produce lift forces of greater magnitude also provides the possibility to utilise new methods of aerodynamics-based orbit and attitude control.

\subsection{On-Orbit Investigations of Gas-Surface Interactions}
A number of investigations of material GSI performance and surface accommodation in the FMF regime have been performed using direct on-orbit measurements and ground-based observations of spacecraft. Review and comparison of studies in this area have been provided by \citet{Moe1998,Moe2011}.

Direct measurement of the remission angle of scattered AO from a vitreous carbon surface was studied on the STS-8 Space Shuttle flight. The diffuse remission spectrum observed, approaching a cosine distribution, indicated that almost full accommodation was occurring at the surface \cite{Gregory1987}. Investigation of scattering angle from an oxidised aluminium surface has also been noted as part of a larger study on erosion characteristics of scattered AO which was conducted on MISSE-FF (Materials International Space Station Experiment Flight Facility) by \citet{Banks2006,Banks2009}. Aerodynamic coefficients resulting from the summed effect of GSI over a spacecraft body have also been studied. For example, the aerodynamic coefficients of the Space Shuttle were measured using accelerometer data during the transitional re-entry phase \cite{Blanchard1986,Blanchard1995}.

Other studies have used observational methods to determine the aerodynamic coefficients of different spacecraft or materials from the attitude motion or orbital trajectory. GSI and surface accommodation can subsequently be investigated by considering the spacecraft attitude and geometry and through comparison to different models. These studies have notably included Paddlewheel \cite{Moe1966} and spherical \cite{Bowman2005,Moe2005a,Pardini2006,Pilinski2011a} satellites, but have also included more complex geometries \cite{Ching1977,Pardini2010} and predictions for time-varying attitude where observed or measured data was not available \cite{Macario-Rojas2018}. However, in the absence of measured data, the results obtained using these methods are typically dependent on modelled atmospheric densities and are therefore subject to their inherent biases and uncertainties \cite{Doornbos2006}. Furthermore, as some of the analysed spacecraft may also have been used during the development and calibration of the density models, some circular logic may be present \cite{Doornbos2012}.

There remains both a lack of knowledge of the physical mechanisms that control GSI behaviour in VLEO and how these apply to different materials and their interactions in the true orbit environment. This is further exhibited by the abundance of GSI models, but lack of consensus on their suitability and application for different materials, surface treatment, altitude range, and period of the solar cycle \cite{Mehta2017}.

The unique combination of a test satellite (SOAR) and an experimental ground facility (ROAR) aims to improve the knowledge of GSIs and the underlying physical mechanisms, leading to improved modelling of aerodynamic forces in VLEO. A systematic investigation of different materials will also seek to identify those that can provide improved aerodynamic performance through specular reflection properties and have atomic oxygen erosion resistance, enabling a new class of spacecraft that can operate sustainably at lower orbital altitudes.

\section{Satellite Design}
The principal scientific objective of SOAR is to investigate the variation of the aerodynamic coefficients of different materials and surface finish at different incidence angle to the oncoming flow and at different orbital altitudes. In-situ measurement of the incident flow environment will be used in addition to measured attitude and orbital parameters to determine the forces and torques experienced by the body. By providing in-situ density measurements of the oncoming flow which can be used directly in the recovery of the fitted aerodynamic coefficients and associated accommodation coefficients, this experimental methodology presents a significant advantage over previous observation-based studies.

\begin{figure*}[t]
	\begin{subfigure}{0.5\linewidth}
	\centering
	\includegraphics[height=70mm,trim={0 0 0 0},clip]{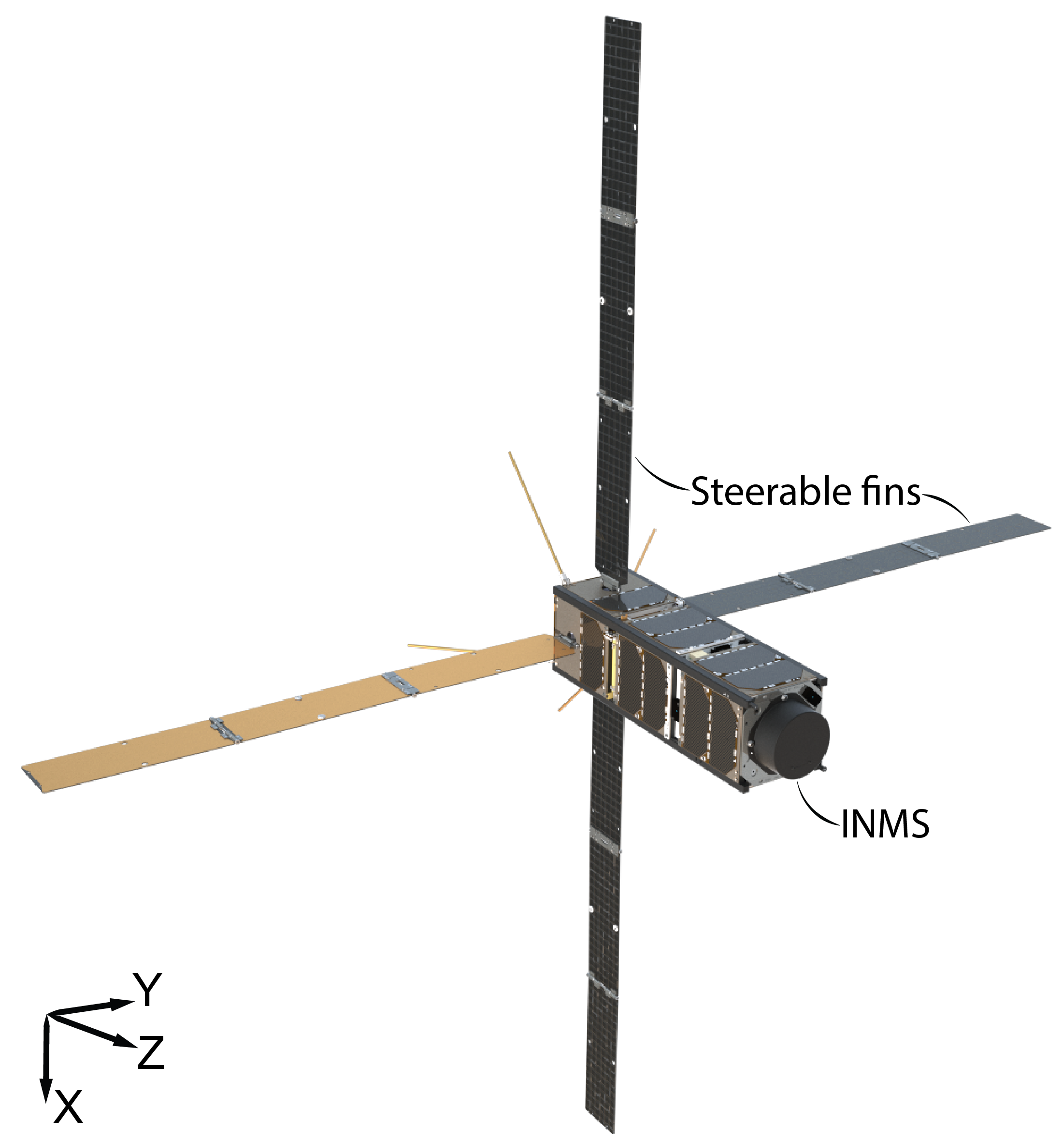}
	\caption{Minimum drag configuration} \label{F:SOARMinDrag}
	\end{subfigure}
	\begin{subfigure}{0.5\linewidth}
	\centering
	\includegraphics[height=70mm,trim={360px 0 500px 0},clip]{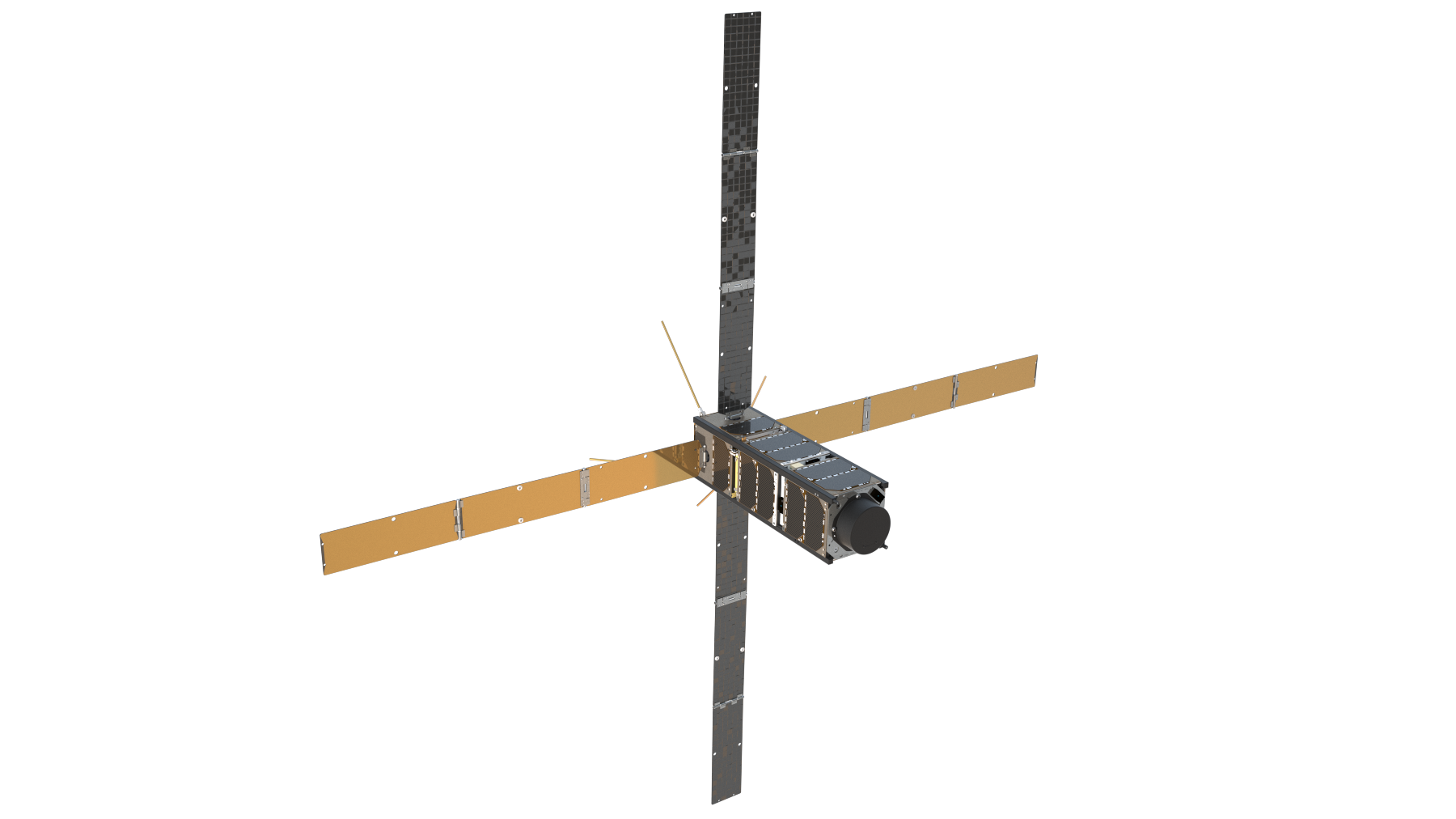}
	\caption{Maximum drag configuration} \label{F:SOARMaxDrag}
	\end{subfigure}
	\caption{Design of the Satellite for Orbital Aerodynamics Research (SOAR) with forward-facing ion and neutral mass spectrometer (INMS) and the steerable fins oriented in the two nominal aerostable configurations.}
	\label{F:SOARGeometry}
	\end{figure*}

SOAR takes the form of a 3U CubeSat developed from the $\Delta$Dsat design of \citet{VirgiliLlop2013a}, previously proposed for the QB50 programme for lower thermospheric exploration and research. The basic geometry of SOAR is shown in \cref{F:SOARGeometry}.

A set of four panels that unfold after launch and deployment into orbit to extend away from the satellite body and can be rotated with respect to the satellite body (and the oncoming flow) have been designed to achieve proposed investigation of material aerodynamic coefficients and to act as aerodynamic control surfaces. These appendages are termed steerable fins herein.

\begin{figure*}[t]
	\centering
	\begin{subfigure}{0.45\linewidth}
	\centering
	\includegraphics[height=70mm,trim={0 0 200px 0},clip]{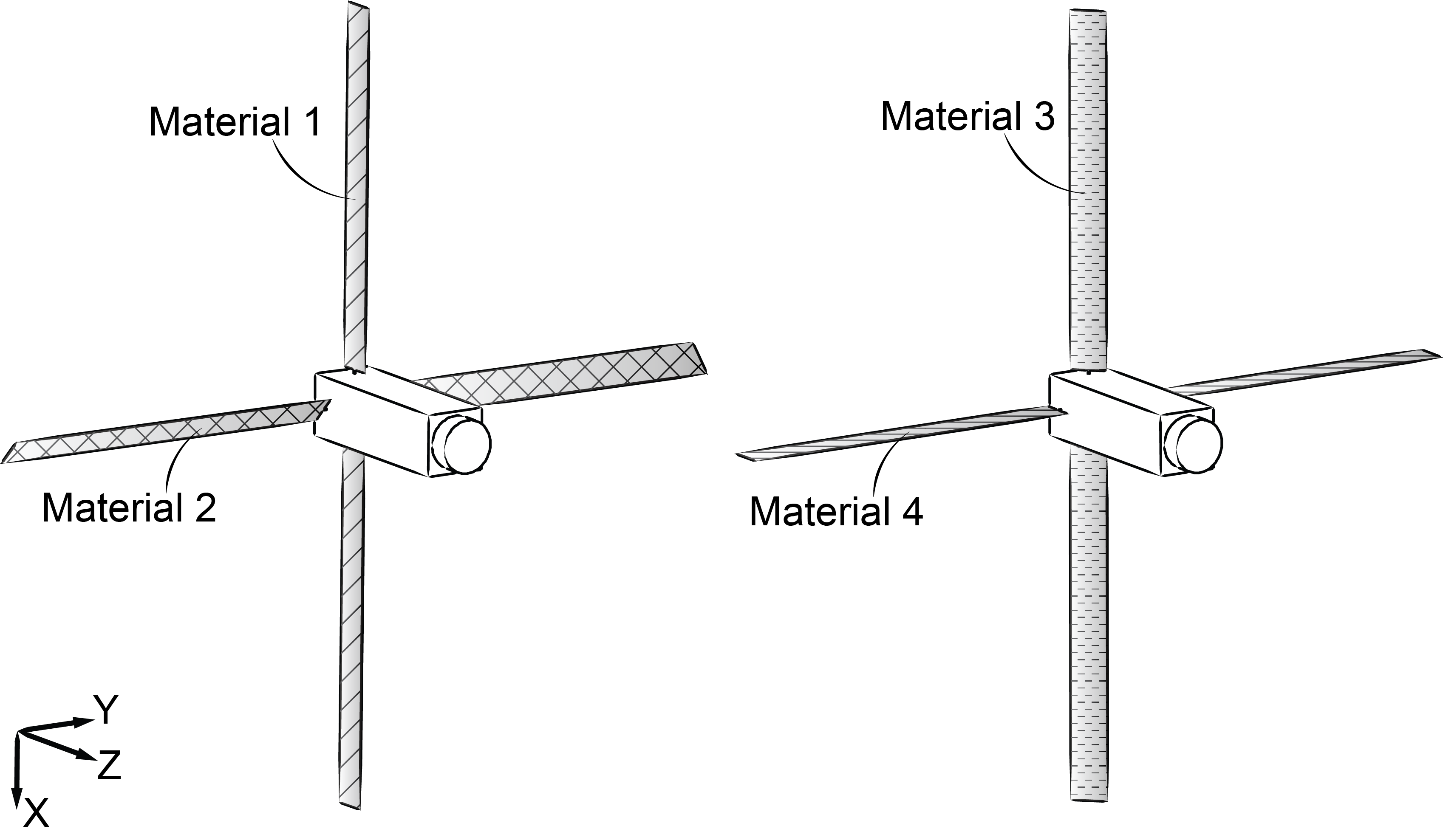}
	\caption{Counter-rotated configuration of the lateral fins.} \label{F:SOARCounter}
	\end{subfigure}
	\begin{subfigure}{0.45\linewidth}
	\centering
	\includegraphics[height=70mm,trim={200px 0 0 0},clip]{SOAR_Configs}
	\caption{Co-rotated configuration of the vertical fins.} \label{F:SOARCo}
	\end{subfigure}
	\caption{Principal experimental configurations of the steerable fins on SOAR showing the corresponding arrangement of the four different test materials.}
	\label{F:SOARConfigurations}
	\end{figure*}

The surfaces of these steerable fins have been coated with four different material coatings with the configuration of similar materials placed on opposing surfaces as indicated in \cref{F:SOARConfigurations}. Through coordinated rotations of the steerable fin, each material can therefore be individually exposed into the flow at varying angles of incidence (neglecting the body of the spacecraft and parallel surfaces).

As described by \citet{VirgiliLlop2013a}, the steerable fins can be operated in pairs in two principal ways; co-rotation and counter-rotation. From the minimum drag configuration and under stable flow-pointing conditions, co-rotation of a single opposing-pair of the steerable fins (see \cref{F:SOARCo}) exposes a single material to the flow and will generate a net lift or side force and therefore a torque (in yaw for the vertical fins or pitch for the lateral fins). The spacecraft will therefore rotate to fly at an angle to the flow. Contrastingly, counter-rotation of a pair of opposing fins (see \cref{F:SOARCounter}) can similarly expose a single material to the flow, but creates opposing lift forces from each steerable fin, resulting in no net side-force and a rolling moment that causes the spacecraft to spin up about the flow-pointing direction.

In order to provide in-situ information about the flow conditions, including thermospheric winds, the spacecraft features a forward-facing ion and neutral mass spectrometer (INMS), labelled in \cref{F:SOARGeometry}. This sensor, improved since the development of the QB50 satellites, includes new time-of-flight (ToF) capability, enabling assessment of the incoming flow velocity in addition to the total atmospheric density and flow composition. To maintain accuracy of the INMS instrument, the spacecraft must be pointed in the direction of the oncoming flow within a given angular range (see \cref{T:Sensors}). Simply, this requires that the spacecraft nominally flies in an attitude that is closely aligned with the direction of the flow.

Attitude control of the spacecraft is principally enabled by a three-axis reaction wheel assembly (tetrahedral configuration of four wheels). A three-axis magnetorquer is also included to perform initial detumbling operations following launch and to enable desaturation and momentum management of the reaction wheels.

Attitude determination for SOAR is provided by fine sun sensors, a magnetometer, and a high-performance IMU (Epson M-G370). Using a unscented Kalman filter (UKF), the combined sensor set is expected to provide an attitude knowledge with an expected uncertainty of less than \ang{1} (3-sigma) even during eclipse. This exceeds the attitude knowledge performance of the antecedent GOMX-3 satellite \cite{Gerhardt2016} and approaches that of the GOMX-4B \cite{Holst2018} satellite, despite not having a star tracker, principally as a result of the improved gyroscope (IMU) performance.

A NovAtel OEM719 GPS receiver provides the precise position ($<$\SI{1.5}{\meter}) and velocity ($<$\SI{0.03}{\meter\per\second}) of the spacecraft and removes dependency of the experiment on ground-based observational tracking information. The accuracy and performance of such miniature commercial-off-the-shelf (COTS) GPS receivers in LEO has been discussed \cite{Montenbruck2006,Montenbruck2008} and demonstrated in orbit, for example on the PROBA-2 \cite{Montenbruck2012} and CASSIOPE satellites \cite{Montenbruck2019}. 

Further parameters of interest relating to the spacecraft design are summarised in \cref{T:SpacecraftParameters}.

\begin{table}
	\caption{Principal geometric and system parameters of SOAR.} \label{T:SpacecraftParameters}
	\centering
	\newcolumntype{L}[0]{>{\arraybackslash}p{0.6\linewidth}}
	\newcolumntype{R}[0]{>{\raggedleft\arraybackslash}p{0.3\linewidth}}
	\begin{tabular}{LR}
	\toprule
	\bf{Property} & \bf{Value} \\
	\midrule
	Mass [\si{\kilo\gram}] & \num{2.88} \\
	Length (in z-axis) ($L_z$) [\si{\meter}] & \num{0.366} \\
	Total Surface Area ($A_T$) [\si{\meter\squared}] & \num{0.225} \\
	CoM (in z-axis from rear) [\si{\meter}] & \num{0.161} \\
	Principal MoIs $ \left\{ \begin{array}{l} x \\ y \\ z \end{array} \right\}$ [\si{\kilo\gram\meter\squared}] & $ \left\{ \begin{array}{l} 0.0392 \\ 0.0392 \\ 0.0288 \end{array} \right\}$ \\
	Residual Magnetic Dipole [\si{\ampere\meter}] & \num{18e-3} \\
	RW Max Torque [\si{\newton\meter}] & \num{23e-6} \\
	RW Max Ang Momentum [\si{\newton\meter\second}] & \num{1.2e-3} \\
	RW Spin Axis MoI [\si{\kilo\gram\meter\squared}] & \num{694.5e-9} \\
	\bottomrule
	\end{tabular}
	\end{table}

\section{Experimental Methodology}
The primary scientific objective of SOAR is to provide in-space measurements of the GSI characteristics of different materials and surface-coatings in the VLEO environment. The steerable fins of SOAR can be used to expose different materials to the oncoming flow at varying incidence angle and at different altitudes as the orbit of SOAR decays. 

The orbit trajectory and attitude of the spacecraft will vary depending on the configuration of the steerable fins with respect to the oncoming flow. With knowledge of the flow conditions and spacecraft position/orientation over time, the aerodynamic forces and torques experienced by the satellite can be estimated and linked to the GSI characteristics of the different surfaces exposed to the flow.

Reconciliation of the force and moment coefficients with the true nature of the GSI mechanics still requires a model for the exchange of energy and momentum of the gas species with the surface and the associated particle reflection/re-emission pattern. However, experimental determination of the aerodynamic coefficients provides valuable in-situ validation data for the ground-based material experiments, in particular those that are planned for the ROAR Facility.

	\subsection{Drag Force Coefficient}
	A body exposed to an oncoming flow will experience forces of an aerodynamic nature, the magnitude and direction of which will be dependent on the orientation of the body with respect to the direction of the oncoming flow. This force is often decomposed into three mutually perpendicular forces in the body axes (axial, normal, and side) with associated coefficients. Alternatively, the components of the force and coefficients with respect to the oncoming flow are considered; drag, lift, and a third mutually perpendicular component (often referred to as side-force or sometimes cross-wind). The term lift, will be used herein to describe both force components perpendicular to the drag, allowing commonality in terminology due to the fourth order rotational symmetry of the spacecraft about the z-axis (see \cref{F:SOARGeometry}).

	The force $\bm{F}$ can be associated with the dimensionless force coefficients $\bm{C_F}$ using \cref{E:AerodynamicForce}, which also expresses the accelerations $\bm{\ddot{x}}$ as a function of the dynamic pressure of the surrounding flow and the spacecraft geometry.

	\begin{equation} \label{E:AerodynamicForce}
	\bm{F} =  \frac{1}{2} \rho v_{\mathrm{rel}}^2 A_{\mathrm{ref}} \bm{C_F} = m \bm{\ddot{x}}
	\end{equation}
	where $\rho$ is the local atmospheric density, $v_{\mathrm{rel}}$ the spacecraft velocity relative to the oncoming flow, and $A_{\mathrm{ref}}$ the reference area.

	Investigation of the drag coefficient of different materials exposed to the flow by the steerable fins was proposed by \citet{VirgiliLlop2013a} for the $\Delta$Dsat mission. In this method, opposing steerable fins are counter-rotated, exposing the same material/coating to the oncoming flow, and nominally producing no net lift/side-forces or pitch/yaw torques but only a net torque in roll. Thus, only an increased nominal drag force is generated by the panel area exposed to the flow and the associated drag coefficient can be determined from the variation in the spacecraft trajectory over a period of time using the orbit determination and free-parameter fitting process described later in \cref{S:FreeParFit}.

	On SOAR, both co-rotated and counter-rotated configurations of opposing steerable fins will be considered. Given the configuration of the material coatings shown previously (\cref{F:SOARGeometry}), the steerable fins can be rotated independently to expose a single material (on two opposing fins) into the oncoming flow to investigate the variation in drag coefficient with incidence angle and at different altitudes.

	The drag coefficient for a given orbital altitude and configuration of the steerable fins can subsequently be recovered by considering the produced aerodynamic acceleration of the spacecraft, expressed by \cref{E:AerodynamicForce}. However, it should be noted that the drag coefficient determined by this method is representative of the whole spacecraft in the given configuration and not only the materials exposed to the flow.

	During these experiments, the attitude control actuators (principally the reaction wheels) will be used to maintain the nominal pointing direction of the satellite into the oncoming flow direction. However, as the steerable fins will be displaced from the minimum or maximum drag condition to expose the different materials to the flow, the spacecraft may have reduced aerostability and experience disturbing aerodynamic torques. Furthermore, the external environmental perturbations may not be periodic in nature and will vary in magnitude depending on a number of factors including the orbital position, spacecraft attitude, solar environment, lighting conditions (sunlight/eclipse), and altitude. For different experiments at different altitudes the stability of the spacecraft may only be maintained by the ADCS for a certain period of time before actuator saturation occurs. The magnitude of the aerodynamic forces, spacecraft stability, and the length of the possible experimental period are critical in determining the expected performance of the investigation. These factors are explored and their impact on the experimental performance estimated and discussed in \cref{S:Uncertainty}.

	\subsection{Lift Force Coefficient}
	Aerodynamic torques experienced by the spacecraft can be described by \cref{E:AerodynamicTorque} in which $\bm{C_M}$ is the aerodynamic moment coefficient set (typically roll $C_l$, pitch $C_m$, and yaw $C_n$) and $l_{ref}$ is an additional reference length ($L_z$ is used herein, see \cref{T:SpacecraftParameters}). The torque can also be defined by the force $\bm{F}$ and associated moment arm $\ell$ or the rotational acceleration $\bm{\ddot{\theta}}$ and the corresponding moment of inertia matrix $\bm{I}$ \cite{NASA1971}.

	\begin{equation} \label{E:AerodynamicTorque}
	\bm{T} = \bm{\ell} \times \bm{F} = \frac{1}{2} \rho v_{\mathrm{rel}}^2 A_{\mathrm{ref}} l_{\mathrm{ref}} \bm{C_M} = \bm{I} \bm{\ddot{\theta}}
	\end{equation}

	The aerodynamic moment coefficients of the satellite can be investigated by analysis of the spacecraft attitude response with the steerable fins configured at different incidence angles with respect to the flow. The lift force coefficient of the different materials exposed to the flow can subsequently be recovered from the aerodynamic moment coefficients by considering the spacecraft geometry and angle of incidence of the fins. Experimental determination of the moment coefficients of SOAR can be performed using either counter-rotated or co-rotated steerable fin configurations.

	A counter-rotated configuration of opposing steerable fins can be used to analyse the rolling moment coefficient. The equal but opposing lift forces produced by the opposing counter-rotated fins act as a couple to generate a net rolling torque on the spacecraft. The rolling moment coefficient can therefore be recovered by considering the evolution of the spacecraft attitude in roll. The lift force coefficient of the exposed surfaces can subsequently be determined by decomposing the spacecraft geometry. Assuming that the body of the spacecraft does not contribute any additional meaningful roll torques, the rolling moment coefficient can be recovered by considering the evolution of attitude in the roll-axis of the spacecraft. Free-parameter fitting of the rolling moment coefficient from the attitude evolution of the spacecraft requires an attitude dynamics model including models for the torques which act on the spacecraft. The orbit trajectory and perturbation models used in the drag coefficient analysis are also required to provide the correct spatial and temporal reference for the selected torque models.

	For a co-rotated configuration of opposing steerable fins a pitching or yawing torque will be produced. In the absence of a correcting control torque, this pitch or yaw torque would cause the spacecraft to rotate (and oscillate about) an equilibrium angle to the flow. By considering the measured evolution of attitude in the pitch/yaw-axis of the spacecraft, the pitch or yaw coefficient for a co-rotated configuration without attitude correction can be recovered. However, if the attitude of the spacecraft is perturbed from the flow-pointing condition the accuracy of the INMS will be compromised and uncertainty in the incidence of the steerable panels to the flow will be increased.

	Alternatively, the reaction wheels can be used to counteract the torque produced by the co-rotated steerable fins and thus attempt to maintain a close to flow-pointing attitude of the spacecraft. A true flow-pointing attitude cannot be realised as knowledge of the oncoming flow direction would be required. Under these circumstances, the measured angular momentum in the reaction wheels rather than the motion of the spacecraft body may be used in the free-parameter fitting process to determine the pitch or yaw moment coefficients.

	In a controlled co-rotated configuration the lift force can also be considered directly through coordinated analysis of the orbital trajectory of the spacecraft and simultaneous parameter fitting of the drag coefficient and the lift force coefficient. However, as the lift force of typical materials is a fraction of the drag force (indicated by the difference in magnitude between the lift and drag coefficients of diffuse surfaces in \cref{F:GSI_CD_CL}), the ability to distinguish the lift force from the measured orbital data in the presence of other sources of perturbation and uncertainty is likely to be limited.
	
	The selection of the most suitable method to investigate the lift force coefficient will be dependent on the attitude and stability characteristics of the spacecraft in each configuration and the expected uncertainty which is associated with the different experimental modes and subsequent data processing.

	\subsection{Orbit Determination and Free-Parameter Fitting} \label{S:FreeParFit}
	In order to recover the aerodynamic forces and torques experienced by the satellite, the orbital position and attitude of the spacecraft during an experimental period can be analysed. However, in addition to the aerodynamic forces and torques of interest, the satellite will experience other external perturbations of varying magnitude, for example due to the non-spherical gravitational field of the Earth, solar radiation pressure, and residual magnetic dipole interactions. The aerodynamic forces and torques experienced by the satellite cannot therefore be simply isolated from the measured position and attitude data.
	
	An orbit determination algorithm can be used to determine the drag coefficient as a \emph{free-parameter} (also seen as the \emph{solve-for} parameter) from the measured orbit position data from a given experimental run and associated configuration of the steerable fins \cite{VirgiliLlop2013a}. The same method can be applied to perform combined orbit and attitude determination to fit and recover a moment coefficient of the spacecraft from the measured orbit position, attitude, and environmental data.

	These methods compare the output of model-based simulations (orbit/attitude propagation) to the data measured on-orbit. Using iterative differential correction, \emph{best-fit} aerodynamic coefficient values can be found by a least-squares method that provides convergence between the measured orbit or attitude trajectory of the spacecraft and the mathematical model of the corresponding motion. Uncertainty in the observations can be accounted for by updates to the initial state vector used for the modelled trajectory at each iteration and differences in the sensor performance for different state variables (e.g. position and velocity) using weighting methods.

	The implementation of this non-linear weighted least squares process \cite{Vallado2013} can be briefly summarised:
	\begin{enumerate}
		\item Import state vectors of experimentally measured orbital elements, attitude quaternions, and atmospheric density for each time step.
		\item Initialise numerical orbit propagation method.
		\begin{enumerate}
			\item Select propagation force and torque models.
			\item Initialise environmental models.
			\item Initialise spacecraft geometric models.
			\end{enumerate}
		\item Set initial guess of the free-parameter (e.g. drag coefficient).
		\item Set weighting matrix based on expected uncertainty of measured state vector parameters (from sensor performance).
		\item Begin iterative scheme:
		\begin{enumerate}
			\item Apply small modifications to each initial state vector component (finite- or central-differencing) based on a small percentage of value or as a function of the weighting matrix.
			\item Calculate the orbit trajectory for each variation of the initial state vector using the orbit propagation method.
			\item Form the partial derivative matrix from differences between each propagated state vectors at each time step.
			\item Calculate update to the initial state vector and free-parameter.
			\item Calculate weighted root mean square (RMS) of residuals (between current iteration and measured trajectory) 
			\item Update state vector and free-parameter. Repeat if RMS has not converged.
			\end{enumerate}
		\item If converged, output state vector and free parameter are best-fit for the observed data and the provided mathematical models (propagation method).
		\end{enumerate}

	The accuracy to which the aerodynamic coefficients can be determined by such a method is primarily dependent on the quality of the experimental data that can be obtained during each test-run. To compare two different spacecraft configurations over a given period of time, it is necessary that the measured trajectories (in orbit or attitude) can first be distinguished from each other in the presence of sensor noise and other uncertainties. For a difference in generated force or torque by the spacecraft in two different configurations this therefore imposes a minimum requirement on the position measurement accuracy (using GPS) and ADCS (attitude determination and control system) measurement accuracy.

	The fidelity of the mathematical model used in the orbit determination process is also critical to the orbit determination process and recovery of the free-parameter (force or moment coefficient). In order to provide convergence towards the measured trajectory, it is necessary that the model incorporates the relevant perturbations with their spatial and temporal variations over the duration of the test-run. The selection of necessary perturbations and modelling fidelity are related to the noise in the measured position, velocity, and attitude. Perturbations that would cause variation in the trajectory of the spacecraft of similar or smaller magnitude than the noise in the measured values can be safely neglected, simplifying the form of the mathematical model.

\section{Attitude Stability and Control}
The presence and use of the steerable fins on SOAR produces a number of different forces and torques which need to be carefully considered to ensure stability and pointing accuracy of the spacecraft throughout its lifetime. Interaction of the spacecraft with the residual atmosphere, solar radiation, and the magnetic and non-spherical gravity fields of the Earth must be considered. The ability to control the attitude and stability of the spacecraft using on-board actuators also requires investigation as the experienced torques vary in relative magnitude with decreasing orbital altitude.

The concept of aerostability is employed by SOAR to provide passive pointing towards the oncoming flow direction in orbit. This aerostability is provided by the steerable fins which are located towards the aft of the spacecraft and thus generate a restoring aerodynamic torque in pitch and/or yaw in response to any misalignment of flow direction with the longitudinal axis of the spacecraft. When each steerable fin is oriented parallel to the longitudinal body axis of the spacecraft (\cref{F:SOARMinDrag}) a minimum drag configuration is generated for the nominal spacecraft attitude. Similarly, when the steerable fins are all oriented normal to the spacecraft longitudinal axis (\cref{F:SOARMaxDrag}) the maximum drag configuration is achieved.

The surface coatings applied to the steerable fins represent a range of expected GSI performance from complete energy accommodation and diffuse re-emission to incomplete accommodation and more specular reflection properties. However, as the GSI properties of these surfaces have not been wholly characterised, the true attitude performance and control capability of the satellite are also uncertain. The different materials may therefore result in the production of different forces and torques when exposed to the flow. 

Modelling of the aerodynamic coefficients for SOAR has been performed using \emph{ADBSat} \cite{Mostaza-Prieto2017a}, an analytical panel-method tool that can implement different gas-surface interaction models and features basic shadowing analysis. Using this tool, a database of aerodynamic coefficients can be calculated from a CAD model of the spacecraft for different orientation angles with respect to the flow (angle of attack and sideslip) and different configurations of the steerable fins.

Sentman's model \cite{Sentman1961} for GSIs has been used in all analyses unless otherwise stated. This model assumes a fully diffuse re-emission pattern of particles with a Maxwellian velocity distribution dependent on the thermal energy accommodation coefficient $\alpha$ and surface (wall) temperature $T_w$. A default accommodation coefficient of $\alpha = 1$ and wall temperature $T_w = \SI{300}{\kelvin}$ have been used unless otherwise stated.

Direct simulation Monte Carlo (DSMC), developed by \citet{Bird1994} and applied in a number of different software tools (e.g. DS3V, DAC \cite{LeBeau1999,LeBeau2001}, dsmcFoam+ \cite{White2018}, and PICLas \cite{Munz2014,Fasoulas2019}), could have alternatively been used to calculate the aerodynamic characteristics of the satellite geometry in the VLEO environment. These methods are able to provide increased fidelity of the modelled flow, including features such as intermolecular collisions, chemical reactions, and electromagnetic/electrostatic interactions. Furthermore, in DSMC more fundamental forms of the GSIs are generally implemented rather than analytical expressions that use mean flow conditions and averaged interactions over flat-plate elements. However, whilst these methods can overcome some of the shortfalls of panel methods (for example for complex and concave geometries), they are significantly more computationally intensive, particularly so when the variation of the aerodynamic coefficients with geometric configuration, attitude, and altitude are needed. Thus, as the geometry of SOAR is relatively simple and the flow can be considered rarefied and free-molecular (for altitudes above $\sim$\SI{150}{\kilo\meter}, see \cref{F:Knudsen}), the panel method could be safely applied for the purpose of the analyses herein.

\subsection{Static Stability}
The static stability provided by different configurations of the steerable fins can be investigated by considering the torque generated by the interaction of the spacecraft geometry with the oncoming flow.

\begin{figure}
	\centering
	\includegraphics[width=85mm]{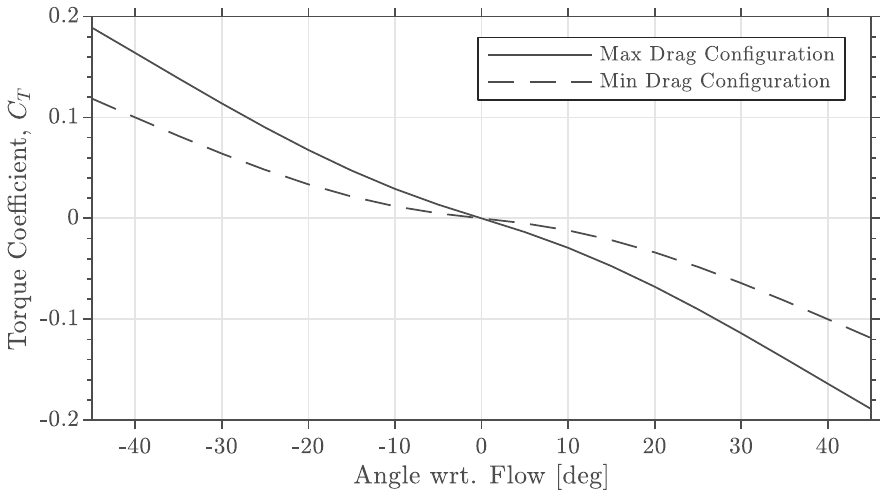}
	\caption{Pitch/Yaw moment coefficient of SOAR with varying angle of incidence with respect to the flow in the minimum (steerable fins parallel to body) and maximum (steerable fins perpendicular to body) drag configurations.}
	\label{F:C_t_MinMaxDrag}
	\end{figure}

The static pitching/yawing moment coefficient of SOAR in the minimum and maximum drag configurations is presented in \cref{F:C_t_MinMaxDrag}. It can be noted that the symmetrical nature of the spacecraft about the roll axis allows for equivalence in pitch and yaw for static analysis. The negative slope of the pitch/yaw moment coefficient with angle of incidence (angle of attack or sideslip respectively) indicates the aerostable nature of these configurations.

\begin{figure}
	\centering
	\includegraphics[width=85mm]{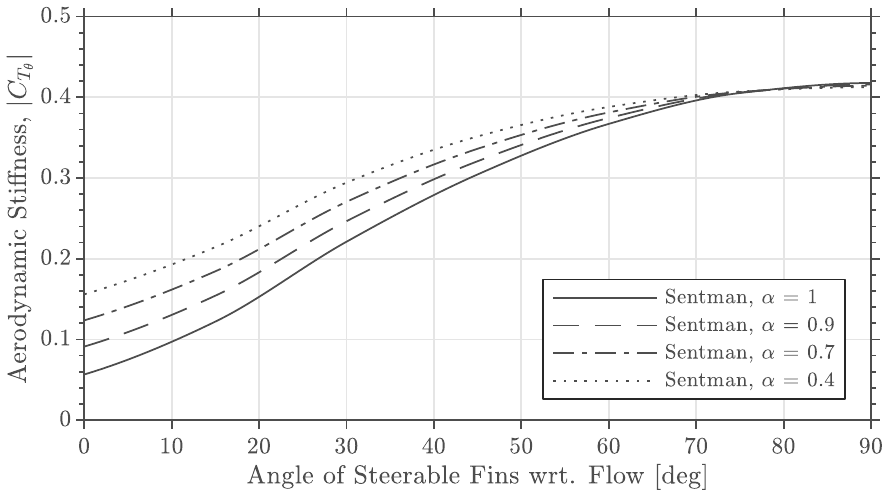}
	\caption{Aerodynamic stiffness (static stability derivative) of SOAR for varying steerable fin angle with respect to the flow.}
	\label{F:C_t_theta}
	\end{figure}

The concept of aerodynamic stability derivatives, or aerodynamic stiffness, for spacecraft at orbital altitudes, can be used to further investigate the expected attitude behaviour for varying geometry and flight conditions \cite{Mostaza-Prieto2016}. The static pitch/yaw stability derivative $C_{T_\theta}$, can be calculated from the gradient of $C_T$ over a small range about the nominal attitude ($\theta = 0$). The variation in static pitch/yaw stability derivative for steerable fin angles over the range of minimum to maximum drag configurations is shown in \cref{F:C_t_theta}. The increase in stability derivative with increasing incidence angle demonstrates that a greater static stability is achieved when a larger panel area is presented to the flow.

The effect of thermal accommodation coefficient on static stability is also shown in \cref{F:C_t_theta}, indicating that aerodynamic stiffness is shown to increase with decreasing thermal accommodation under the assumptions of Sentman's GSI model. This effect is more marked at shallow steerable fin incidence angle with respect to the flow.

In order to characterise the performance of different materials and surface-coatings in orbit, during the experimental periods only one pair of steerable fins will be rotated with respect to the flow at any given time. With this configuration a total of four materials or surfaces can be characterised during the mission, two per pair of opposing steerable fins.

\begin{figure}
	\centering
	\includegraphics[width=85mm]{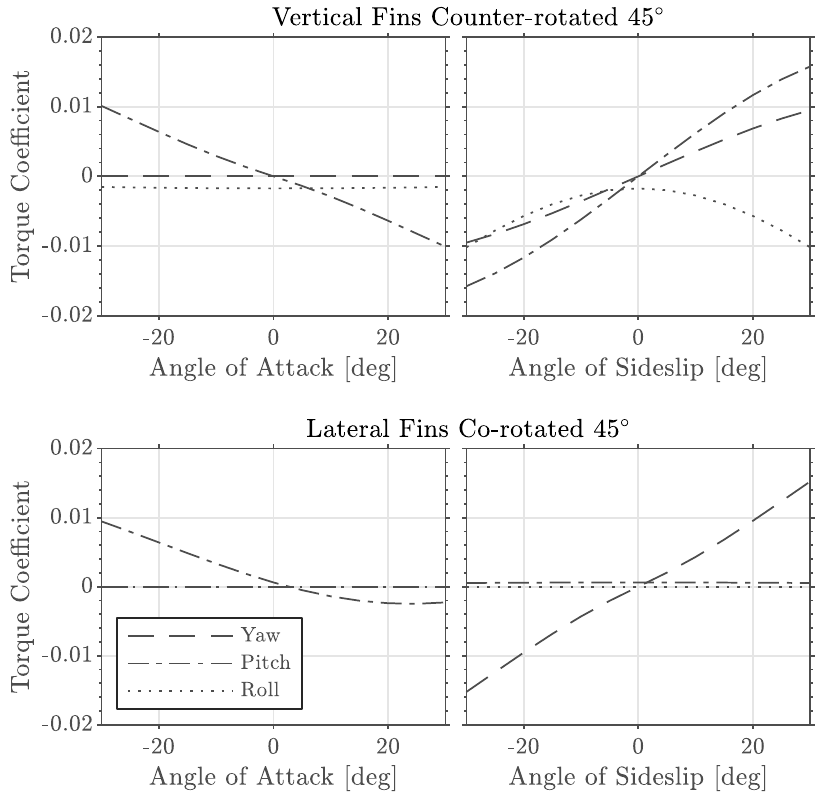}
	\caption{Aerodynamic torque coefficient response for SOAR with different counter-rotated or co-rotated steerable fin configuration.}
	\label{F:SOAR_Stability}
	\end{figure}

When a single pair of steerable fins is counter-rotated with the spacecraft pointing into the direction of the oncoming flow a net rolling torque is generated but no net pitch or yaw torques are created. However, if the relative direction of the flow changes (for example due to atmospheric co-rotation or thermospheric winds) or the attitude of the satellite is perturbed, induced torques are generated due to a variation in the projected area of the counter-rotated fins to the flow. Under these conditions a net pitching torque arises due to the difference in area presented to the flow by the rotated fins. The plot of torques for counter-rotated vertical fins is shown in \cref{F:SOAR_Stability} (top), demonstrating that induced torques in pitch due to angle of sideslip have a positive gradient about the equilibrium, and are therefore disturbing rather than restoring. Furthermore, these torques grow at a faster rate than the restoring torques generated by the lateral fins due to the change in pitch angle. If the flow is therefore offset with respect to the spacecraft body in yaw the spacecraft responds with coupled motion in the pitch axis as a result of the set angle of the steerable fins. Equivalent behaviour is demonstrated for counter-rotated lateral fins and an offset in angle of attack. This effect is termed pitch-yaw coupling henceforth.

Co-rotation of a pair of opposing steerable fins generates a net torque in pitch or yaw, but no net torque in roll. \cref{F:SOAR_Stability} (bottom) shows the torques in pitch and yaw for a configuration in which the lateral fins are co-rotated, demonstrating a small bias in pitch torque when the spacecraft is aligned with the direction of the oncoming flow. However, as the pitch (angle of attack) is increased by a small amount ($\sim$\ang{3}) the pitch torque crosses zero with a negative gradient. The spacecraft therefore demonstrates stability in pitch at this equilibrium angle with respect to the flow. The spacecraft is also shown to be stable in yaw about the oncoming flow direction.

\subsection{Dynamic Stability} \label{S:Dynamics}
In order to understand the evolution of attitude over time and in the presence of perturbing torques the dynamic response of the spacecraft must be considered. The orbital and attitude response of SOAR in the VLEO environment can be investigated using a 6-DOF simultaneous orbit and attitude propagation method. This method is based on the numerical solution of the complete equations of motion for an orbiting satellite (kinematic and dynamic motion \cite{LandisMarkley2014}) for which varying force and torque model inputs can be provided. In these simulations the forces and torques associated with the Earth gravitational potential (EGM96/2008 \cite{Lemoine1998,Pavlis2012}), solar radiation pressure, residual magnetic dipole interactions (IGRF-11 {\cite{Finlay2010}}), varying atmospheric density (NRLMSISE-00 \cite{Picone2002}), and thermospheric winds (HWM93/07 \cite{Hedin1996,Drob2008}) are implemented. It is important to recall that the results presented herein remain subject to the assumptions and limitations of the implemented GSI model and the input parameters used and therefore may still differ substantially from the true behaviour in orbit.

For rotationally symmetric configurations the previous analyses showed that aerostability ensures that restoring torques will be produced in response to changes in the oncoming flow direction. However, due to the FMF nature of the surrounding atmospheric environment, natural damping of any generated angular velocity is not generated. Therefore, given an external perturbation and without any additional damping input, the spacecraft will begin to oscillate. The frequency of this oscillation is dependent on the initial disturbance, stability derivative, and the environmental conditions \cite{Mostaza-Prieto2016}.

\begin{figure*}
	\centering
	\includegraphics[width=180mm]{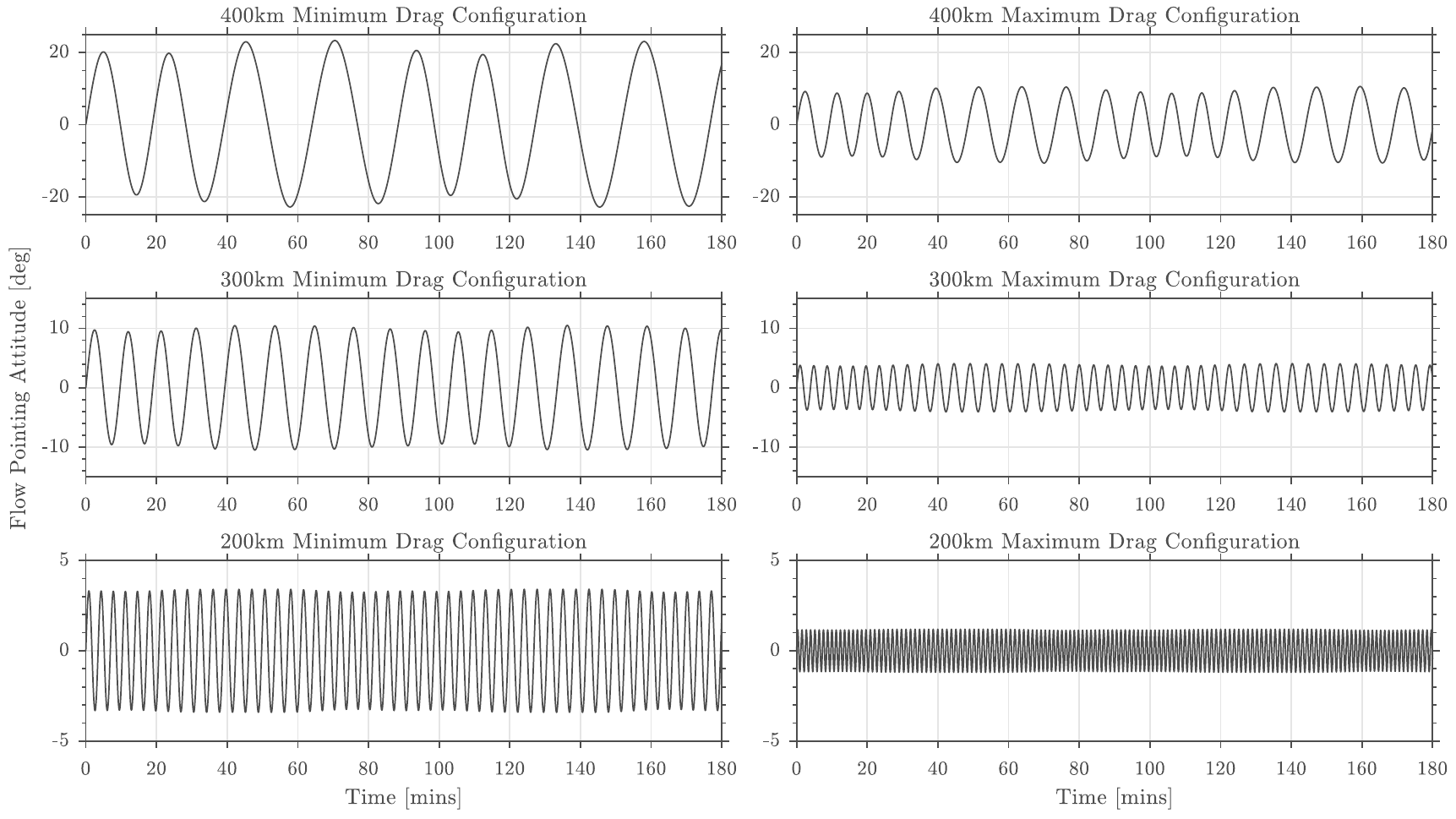}
	\caption{Nominal uncontrolled pitch response of SOAR for varying altitude and minimum/maximum drag fin configuration in an equatorial orbit. Only spherical gravity and aerodynamic torques are considered in an equatorial orbit. An initial pitching rate of \SI{0.1}{\degree\per\second} is applied.}
	\label{F:Alt_Min_Max}
	\end{figure*}

The nominal response of SOAR in the minimum and maximum drag configuration for varying orbital altitude and in the absence of further perturbing torques is presented in \cref{F:Alt_Min_Max}. The responses demonstrate the basic aerostable nature of the spacecraft in the maximum and minimum drag configurations and that the oscillatory amplitude decreases and frequency increases with increasing aerodynamic stiffness and dynamic pressure.

The oscillatory modes can be internally damped using attitude control actuators as discussed by \citet{VirgiliLlop2014}, significantly reducing the range over which the attitude of the spacecraft varies. However, due to the presence of further perturbing torques in the real orbital environment (e.g. atmospheric co-rotation, thermospheric winds, solar-radiation pressure, gravity gradient), and errors or incompatibilities associated with real attitude actuator systems (e.g. magnetorquer availability and cross-coupling), the true dynamic response is more complex.

Methods of control for aerostable spacecraft have been presented in the literature. \citet{Psiaki2004} presents a compass-like PID control method which utilises magnetorquers to provide three-axis stabilised nadir-pointing capability to a 1U CubeSat with a shuttlecock configuration of deployed ``feathers''. \citet{Auret2011} subsequently applied this method to a 3U CubeSat geometry which included a pair of actuating ``paddles'' that provide control capability about the roll axis. Aerostability with assisted damping has also been successfully demonstrated in orbit. The DS-MO spacecraft featured extended aerodynamic skirts and a gyrodamper mechanism \cite{Sarychev2007}. The Passive Aerodynamically-stabilized Magnetically-damped Satellite (PAMS) had a cylindrical geometry with a biased centre-of-mass that provided aerostability and magnetic hysteresis rods to provide damping \cite{Kumar1995}. The Gravity field and steady-state Ocean Circulation Explorer (GOCE) spacecraft featured rear-mounted aerodynamic fins and was equipped with magnetorquers for damping \cite{Drinkwater2007}. MagSat also demonstrated trim in pitch using an aerodynamic boom of variable length \cite{Stengle1980,Tossman1980}. Three-axis aerodynamic pointing control has also been considered for a ``shuttlecock'' geometry \cite{Gargasz2007} and feathered configurations \cite{VirgiliLlop2014} similar to SOAR.

A significant challenge in the control of an aerodynamic spacecraft is that the true oncoming flow vector is generally not known a priori. A true reference vector for three-axis flow-pointing control is therefore missing. A reference vector including the effect of atmospheric co-rotation can be provided, however the prediction of thermospheric winds using available models is associated with much greater uncertainty. Alternatively, a sensor which can provide the oncoming flow vector could be used to provide in-situ measurements for active control methods. However, proposals for such instruments for use in VLEO are only just emerging \cite{Eberhart2015,Verker2020}.

SOAR will be launched to the International Space Station and deployed into an \ang{51.6} inclination orbit with an initial altitude of approximately \SI{400}{\kilo\meter}. Whilst forecasts for solar cycle 25 vary \cite{Hathaway2016,Cameron2016,Bhowmik2018} the deployment will occur during a period of minimum solar activity and the atmospheric density at this altitude will therefore be characteristically low.

\begin{figure}
	\centering
	\includegraphics[width=85mm]{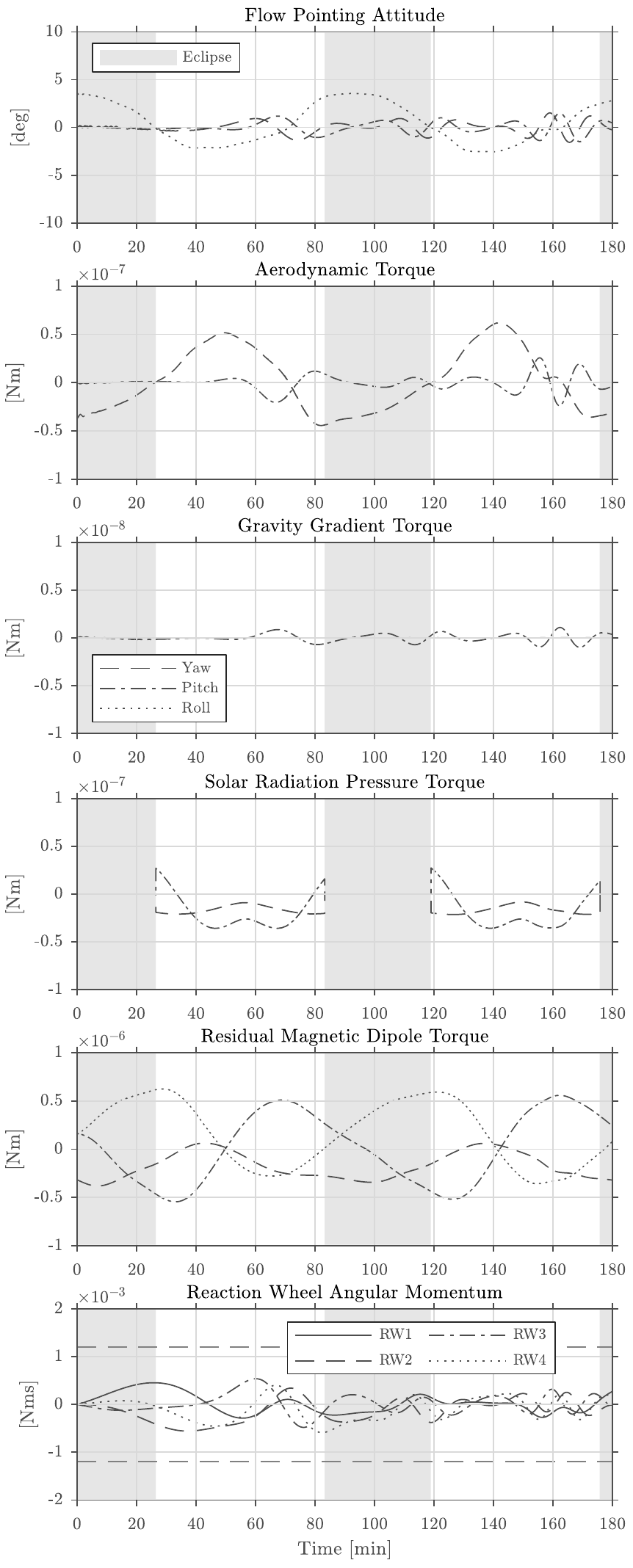}
	\caption{Environmental torques and controlled attitude of SOAR at \SI{400}{\kilo\meter} altitude and \ang{51.6} inclination in maximum drag configuration.}
	\label{F:SOAR_Aerostable_Control_400km}
	\end{figure}

The attitude response of SOAR in the maximum drag configuration with reaction wheel damping and proportional control is shown in \cref{F:SOAR_Aerostable_Control_400km}. The restoring aerodynamic torques experienced by the spacecraft are shown to be of a similar magnitude to the solar radiation pressure torques and over an order of magnitude smaller than the residual magnetic dipole torques. The use of the reaction wheels is therefore important at this altitude to assist the stability of the spacecraft and maintain pointing of the spacecraft close to the oncoming flow direction.

The accumulation of the perturbing torques may also necessitate periodic management of the angular momentum within the reaction wheels to avoid saturation. The typical disadvantages of magnetorquer cross-coupling and limited availability (due to instantaneous orientation of the magnetic field) must be accommodated whilst the power consumption of the actuators must be considered for extended use and through eclipse periods. As SOAR naturally decays to lower altitudes, the magnitude of the aerodynamic torques will increase, and the aerostability of the spacecraft when in the minimum or maximum drag configurations will improve, reducing the requirements on the attitude actuators.

Given the relative magnitude of the different external perturbations, particularly the residual magnetic dipole, it may be more difficult to perform the experiments at higher altitudes where the ``signal'' (aerodynamic forces and torques) is low in comparison to the sources of ``noise'' (other perturbing forces and torques). It may therefore be necessary to allow the satellite to initially decay in altitude before commencing the experimental operations.

\subsection{Experimental Configurations}
During the experimental operations co-rotated or counter-rotated configurations of the steerable fins will be utilised to expose the different materials to the oncoming flow, modifying the natural stability and therefore attitude dynamics of the spacecraft.

In the counter-rotated configuration, when the spacecraft is nominally pointed towards the oncoming flow direction, no net torques are generated in pitch or yaw. A rolling moment is however produced due to the opposing lift forces generated on the two exposed surfaces, which if uncontrolled will cause the spacecraft to spin up. If the spacecraft is disturbed from its equilibrium flow pointing configuration pitch and/or yaw torques will be produced and the effect of pitch-yaw coupling will act to further disturb the attitude of the spacecraft from the flow-pointing direction.

In a three-axis controlled mode, a maximum duration on operations with fins rotated in a counter-rotated configuration is imposed by the build-up of angular momentum and saturation of reaction wheels. This is a function of the atmospheric density, incidence angle of the steerable fins, and material performance. The thermospheric wind, solar activity, and other external disturbance torques also contribute to this attitude performance, but vary with greater uncertainty. At lower altitudes the time-period over which spacecraft can be operated successfully may therefore be significantly limited for some counter-rotated configurations, the impact of which will be discussed later in \cref{S:Uncertainty} with regards to the experimental uncertainty.

In alternative co-rotated fin configurations, net pitch or yaw torques are generated by the common incidence angle of the two opposing steerable fins. This causes the spacecraft to rotate and fly at an angle to the oncoming flow with an associated oscillatory motion about the new offset equilibrium attitude that results from the aerostability of the spacecraft. The control actuators may be used to correct the nominal pointing direction of the satellite such that the INMS will be realigned close to the flow, ensuring the accuracy of the measured density and flow velocity information. However, as a bias in the pitch or yaw torques exists, angular momentum will again accumulate in the reaction wheels, eventually causing saturation of the attitude control system.

Contrastingly, for the investigation of lift force coefficient, the roll axis can be left uncontrolled allowing the net torque to accumulate and the acceleration and angle in roll to be measured by the ADCS. Control of the pitch and yaw axes are maintained to keep the spacecraft pointing close to the oncoming flow direction.

\section{Experimental Performance} \label{S:Uncertainty}
The expected performance of the mission and the experimental determination of the aerodynamic coefficients can be investigated by testing the free-parameter fitting and least-squares processes using simulated orbit and attitude data. This data is modified with noise to represent the expected in-orbit sensor performance. A Monte Carlo approach is also implemented to encompass variation in the initial conditions (epoch, orbit, and attitude) and to vary the addition of noise to the data for each simulated run.

The free-parameter fitting process utilises a least-squares orbit determination algorithm, described in \cref{S:FreeParFit}. This process seeks to minimise the error between the reference (simulated or on-orbit) data and mathematically modelled data by varying the \emph{free} values of the aerodynamic coefficients in the presence of known measurement uncertainties. The process is iterative and is terminated by convergence criteria based on the weighted RMS of the residuals between the reference and modelled data.  Central-differencing methods are used to account for the errors in the initial condition of the state vector due to measurement uncertainties.

Simulated on-orbit data is first produced using the high-fidelity attitude and orbit propagation method described in \cref{S:Dynamics}. Orbit and attitude noise is produced by considering the performance parameters of the GPS and ADCS sensors, reported in \cref{T:Sensors}. Uncertainty on the angular velocity of the reaction wheels and angular position of the steerable fins has also been similarly introduced. The expected in-orbit performance of the INMS is also simulated. The measured density is first produced using the NRLMSISE-00 \cite{Picone2002} atmosphere model, informed by orbit and attitude data, and modified for GPS and ADCS sensor and acquisition errors and noise. This density is then subsequently transformed using the INMS instrument uncertainty and angular dependency, expressed as full-width half-maximum (FWHM) measures in \cref{T:Sensors}. 

This modified data is subsequently used to perform the orbit and attitude determination processes. For recovery of the drag coefficient only orbit propagation is performed, whilst for the moment coefficient combined attitude and orbit propagation is performed. In both cases, the initial ``measured'' state vector is used to set-up the propagation method. The force and torque models used in the propagation are modified to incorporate further uncertainty and to represent the reduction in or lack of knowledge of the true in-orbit environment and spacecraft interactions. For the drag coefficient fitting, the the thermospheric winds are neglected, whilst for the moment coefficient cases the residual magnetic dipole interactions are also excluded from the attitude propagation scheme.  

The expected performance of the experiments at different orbital altitudes and steerable fin configurations can be obtained by considering the standard deviation of the returned aerodynamic coefficient after a number of Monte Carlo iterations. However, given that such a Monte Carlo simulation only provides a random sample of results, the confidence of the standard deviation should also be considered, within which the population standard deviation would be expected to lie with the given confidence.

Reducing the width of the standard deviation increases the resolution of the experiment and can be most simply achieved by increasing the signal-to-noise ratio of the experiment. For the on-orbit experiments this may be achieved by increasing the test run time or the magnitude of the force or torque to be measured. However, in many cases these parameters are restricted by the platform design (e.g. steerable fin size, reaction wheel capability) or on-orbit operations (e.g. power and downlink budget). Reducing the uncertainties associated with the measured on-orbit data may also be helpful, for example improved in-situ density measurements and position/attitude knowledge. Finally, improved knowledge and modelling of the unmeasured disturbances or perturbations would further reduce the uncertainties within the free-parameter fitting and orbit determination process, primarily requiring characterisation of the residual magnetic dipole and solar radiation pressure interactions of the satellite.

\begin{table}
	\caption{Summary of expected satellite sensor performance.} \label{T:Sensors}
	\centering
	\begin{tabular}{lr}
	\toprule
	\bf{Instrument} & \bf{Uncertainty} \\
	\midrule
	GPS Position [\si{\meter}] & \num{2.5}  ($1\sigma$) \\
	GPS Velocity [\si{\meter\per\second}] & \num{45e-3} ($1\sigma$) \\
	ADCS Angle [\si{\radian}] & \num{0.2e-3} ($1\sigma$) \\
	ADCS Angular Velocity [\si{\radian\per\second}] & \num{25e-3} ($1\sigma$) \\
	INMS Number Density [\si{\per\cm\cubed}] & $\sqrt{N} + 0.7$ ($1\sigma$) \\
	INMS Horizontal Acceptance [\si{\radian}] & 0.279 (FWHM) \\
	INMS Vertical Acceptance [\si{\radian}] & 0.035 (FWHM) \\
	Steerable Fin Rotation Angle [\si{\radian}] & \num{0.015} ($1\sigma$) \\
	\bottomrule
	\end{tabular}
	\end{table}

\subsection{Drag Coefficient}
\begin{figure*}[tb]
	\centering
	\includegraphics[width=180mm]{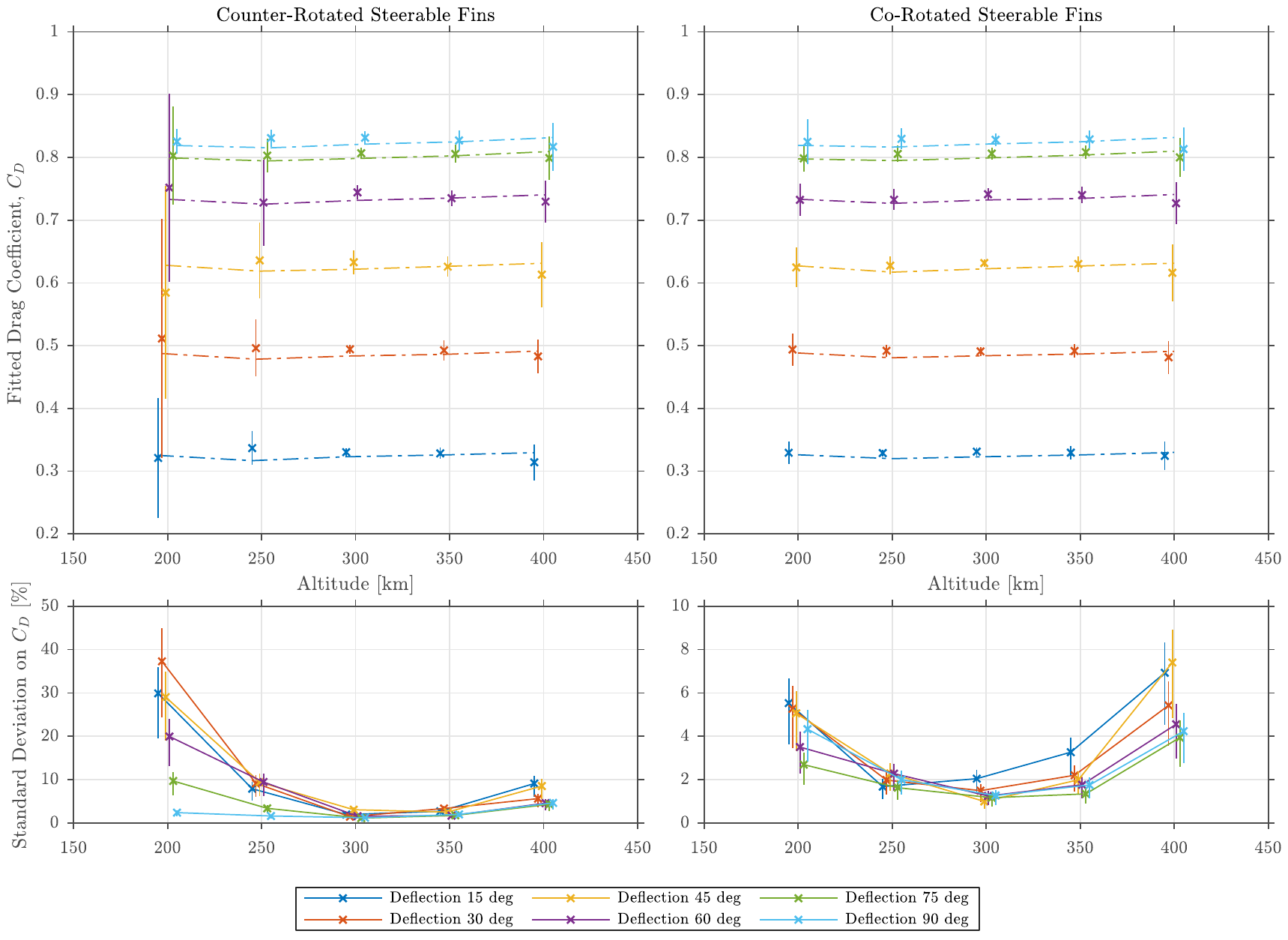}
	\caption{Drag coefficient determination performance for counter-rotated and co-rotated steerable fin configurations. Sample mean fitted drag coefficient (top), referred to reference area $A_T/2$, is given top with error-bars representing the associated standard deviation. Reference lines indicate the modelled GSI value. The standard deviation is given bottom with error bars representing the 95\% confidence interval. Data points have been shifted slightly in the x-axis to allow for visibility of overlapping error bars.}
	\label{F:Exp_Drag_Counter_Co_3Axis}
	\end{figure*}

The returned drag coefficients from this data reduction process are shown in \cref{F:Exp_Drag_Counter_Co_3Axis} for counter-rotated and co-rotated steerable fin configurations over a range of different incidence angles and altitudes. A duration of \SI{120}{\minute} is targeted, a limit imposed by the expected power balance achievable by spacecraft in the experimental mode. Three-axis reaction wheel control is implemented to maintain an approximately flow-pointing attitude and stability. The test-run is aborted if the angular range in pitch or yaw exceeds the INMS instrument acceptance limit or the reaction wheels approach saturation. Dashed lines for the underlying GSI model-based drag coefficient (calculated using Sentman's model with $\sigma = 1$ and the ADBSAT tool) at each altitude and configuration are provided as a reference. It should be noted that under the assumption of diffuse re-emission properties, the counted-rotated and co-rotated configurations an incidence of \ang{90} are equivalent. The very minor differences in the presented results are therefore due to variations that arise from the implementation of the Monte Carlo simulation.

The results for the counter-rotated and co-rotated configurations are very similar for altitudes between \SIrange{400}{300}{\kilo\meter}. However, at altitudes at and below \SI{250}{\kilo\meter} the results from the co-rotated configuration demonstrate much smaller standard deviations and mean values closer to the reference values. This is principally due to the longer experimental periods that can be maintained by the attitude control system in the co-rotated configuration before the reaction wheels saturate, in general more than double than the counter-rotated configuration. More orbital position information is therefore provided against which the least-squares orbit determination process can best fit the experimentally determined drag coefficient, reducing the experimental uncertainty.

Increasing the experimental duration at higher orbital altitudes would similarly assist in improving the experimental uncertainty for both the counter-rotated and co-rotated steerable fin configurations. However, this is challenging due to the power budget of the satellite. The increase in standard deviation at \SI{400}{\kilo\meter} is due to low atmospheric density at higher altitudes and therefore the limited effect that the drag will have on the orbit over a period of only \SI{120}{\minute} in the presence of the GPS position measurements, particularly for shallow steerable fin incidence angles. The greater relative magnitude of the additional disturbing perturbations also contribute to the increased experimental uncertainty at the higher altitudes.

For both the co-rotated and counter-rotated configurations the minimum experimental uncertainty is expected to be achieved at approximately \SI{300}{\kilo\meter} altitude where a balance between the experimental duration, measurable effect on the orbital trajectory, and magnitude of external perturbations is found.

These results indicate that drag coefficients determined from the on-orbit experiments and associated measured data for different steerable fin incidence angles (in \ang{15} increments) are likely to be largely identifiable and distinguishable from one another. In the co-rotated configuration smaller increments in incidence angle may also be discernible, particularly at shallower deflections.

Difference in the experimentally determined results from the reference drag coefficients arises primarily from the variation in the area of the satellite surfaces projected into the flow. This occurs as the satellite attitude varies with respect to the flow as a result of the aerodynamic and other environmental torques. Further, the attitude control system does not have knowledge of the oncoming flow direction and therefore uses the offset LVLH frame as a reference. Additional sources of error can be attributed to the sensor accuracy and noise parameters (INMS, GPS, and attitude) and the modelled forces used in the least-squares orbit determination fitting process. Of the modelled forces, the solar radiation pressure exhibits the greatest uncertainties. Improved knowledge of the solar radiation pressure interaction with the different external surfaces of the satellite could improve the estimation of these effects.

The dependence of the drag coefficient with altitude appears to be less clearly identifiable from the expected experimental performance. This is due to the relatively small variation in drag coefficient which is expected over the available altitude range (\SIrange{200}{400}{\kilo\meter}) in comparison to the experimental uncertainty. However, Sentman's GSI model with a single accommodation coefficient ($\sigma=1$) has been used in this analysis, representing typical diffusely re-emitting materials, for example contaminated metallic surfaces \cite{Moe2011}. When complete accommodation and diffuse re-emission is assumed significant variation in the drag coefficient with altitude is not expected and will only be driven by the variation in speed ratio and thermospheric temperature.

\subsection{Rolling Moment Coefficient}
\begin{figure}[tb]
	\centering
	\includegraphics[width=85mm]{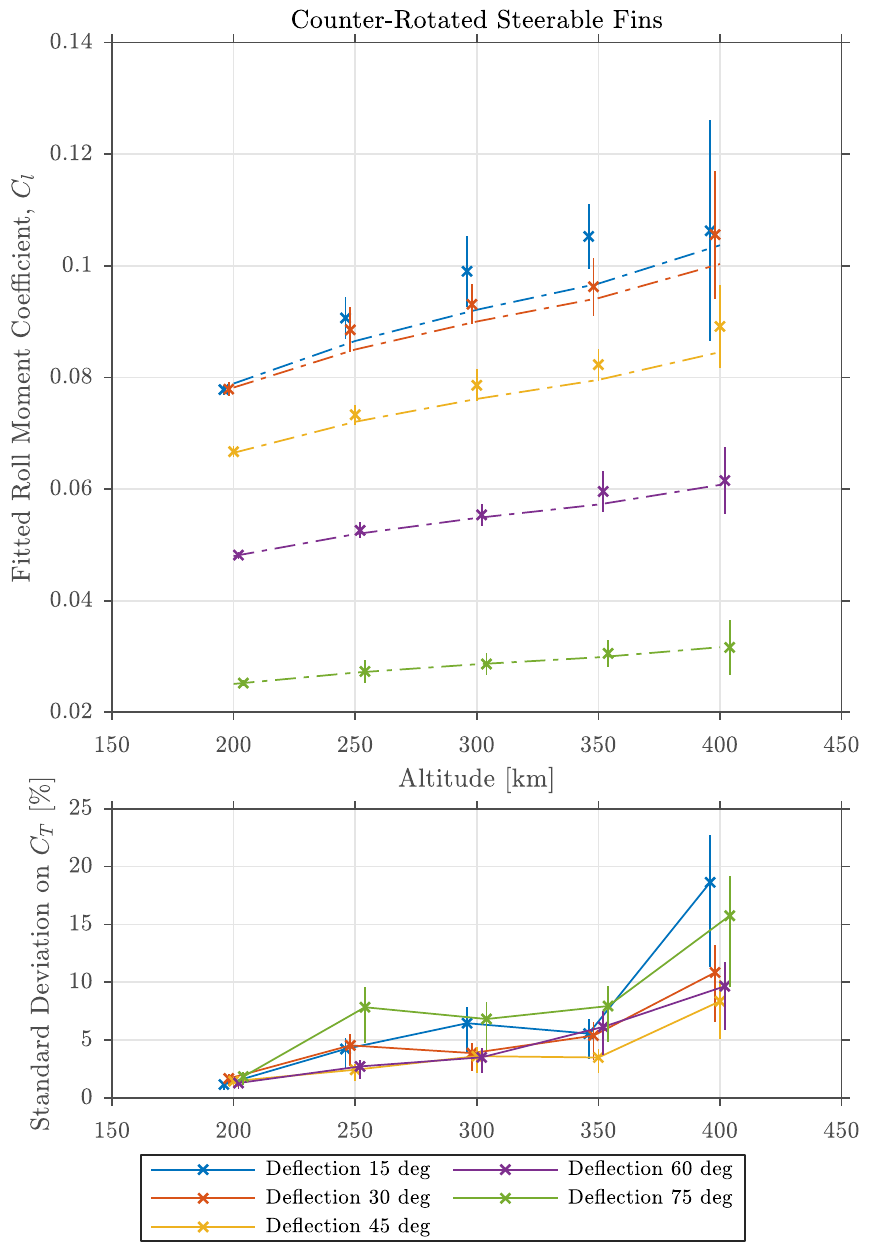}
	\caption{Rolling moment coefficient determination performance for counter-rotated steerable fin configurations. Sample mean fitted torque coefficient (top), referred to reference area $A_T/2$ and reference length $L_z$, is given with error-bars representing the associated standard deviation. Reference lines indicate the modelled GSI value. The standard deviation (bottom) is given with error bars representing the 95\% confidence interval. Data points have been shifted slightly in the x-axis to allow for visibility of overlapping error bars.}
	\label{F:Exp_Lift_Counter_3Axis}
	\end{figure}

The results for the fitted rolling moment coefficient experiments are shown in \cref{F:Exp_Lift_Counter_3Axis} for varying counter-rotated steerable fin incidence angles and altitude under two-axis (pitch and yaw) control.

The experimental uncertainty is generally seen to increase with orbital altitude. This is attributed to the effect of unmeasured and poorly modelled attitude perturbations on the satellite during the experiment. These are most significant when the atmospheric density is lowest and the aerodynamic torques are therefore relatively low in magnitude. Residual magnetic dipole interactions are the most significant of these effects and are not included in the orbit determination algorithm. However, if knowledge or modelling of the time-varying residual magnetic dipole of the satellite can be obtained this may be incorporated into the analysis. Similarly to the drag coefficient experiments, improvement of the knowledge and modelling of solar radiation pressure effects could also be beneficial.

For most steerable fin configurations the rolling moment coefficient is clearly identifiable against the other results. However, at lower incidence angles (\ang{15} and \ang{30}) the error bars overlap indicating that these configurations may not be distinguishable from each other from the on-orbit measurements. At greater incidence angles the variation between smaller increments in steerable fin angle may be possible, particularly at lower altitudes.

In contrast to the drag coefficient results presented previously, the dependence of torque coefficient on orbital altitude appears to be more marked, even for assumed fully accommodated gas-surface interactions. However, this variation may remain obscured by the experimental uncertainties, particularly for larger steerable fin incidence angles (\ang{60} and \ang{75}) that vary more slowly with altitude.

\section{Concluding Remarks}
This paper has described the proposed method for determination of the aerodynamic coefficients of different materials on SOAR, a scientific CubeSat due to be launched in 2021. The presented analysis and simulated dynamics of the SOAR geometry demonstrate the aerostable nature of the design in the nominal maximum and minimum drag modes and the use of the steerable fins in both counter-rotated and co-rotated modes to perform the proposed aerodynamics characterisation experiments.

Using the combination of the INMS and the steerable fin payloads, on-orbit experimental assessment of the aerodynamic coefficients of different materials at varying incidence to the oncoming flow will be performed. These experiments will be repeated as the orbit of SOAR decays to investigate the variation with orbital altitude. The modelled uncertainty of these experiments indicates that the drag and lift coefficients at different incidence be determined from the measured parameters in the presence of the disturbing and perturbing forces and torques present in VLEO. The uncertainty of drag coefficient measurements was shown to be minimised around an altitude of \SI{300}{\kilo\meter}, whilst the lift coefficient experiment generally demonstrates improvement as the altitude is reduced further. These insights will be used to plan the operations of the SOAR mission.

The purpose of this on-orbit experimentation is to provide valuable in-situ validation data for a more extensive investigation of rarefied-flow GSIs to be performed on the ground with the aim to improve knowledge of GSI mechanisms and the associated models that describe this behaviour. A systematic study to identify materials that can increase aerodynamic performance at lower orbital altitudes will also be performed. SOAR will test two such novel materials with promising drag-reducing characteristics in-orbit.

\section*{Acknowledgements}
This project has received funding from the European Union's Horizon 2020 research and innovation programme under grant agreement No 737183. This publication reflects only the view of the authors. The European Commission is not responsible for any use that may be made of the information it contains.

\bibliographystyle{elsarticle-num-names}
\bibliography{SOAR}

\end{document}